\documentstyle[12pt]{article}
\title{Self-gravitating bosons at nonzero temperature}
\author{Neven Bili\'c$^*$  and Hrvoje Nikoli\'c  \\
Theoretical Physics Division, Rudjer Bo\v skovi\'{c} Institute, \\
P.O.B. 180, 10002 Zagreb, Croatia \\
{\normalsize
$^*$bilic@thphys.irb.hr
}}
\begin{document}
\maketitle
\begin{abstract}
A system
of charged bosons at finite temperature and
chemical potential is studied in a general-relativistic
framework.
We assume that the boson fields interact only gravitationally.
At sufficiently low temperature
the system
exists in two phases: the gas and the condensate.
By studying the condensation process numerically
we determine the critical temperature $T_{\rm{c}}$
at which the condensate emerges.
As the temperature decreases, the system
eventually settles down in the ground state
of a cold boson star.
\end{abstract}

\section{Introduction}
\label{intro}
Bose-Einstein condensation is an extensively studied
subject of interest to almost all branch\-es of physics.
It is probably the simplest example of critical phenomena
since it exists theoretically even in noninteracting systems
\cite{kap1,hab1,hab2}.
In an interacting theory containing complex scalar fields,
Bose-Einstein condensation becomes a highly nontrivial
phenomenon.
In self-interacting scalar theories
there are cases, such as $\phi^4$ theory,
where a homogeneous condensate is a stable ground state.
In some cases, the Bose-Einstein condensate at zero temperature
exists in the form of stable nontopological solitons
known as Q-balls
\cite{fri,col}.
Another interesting example is a scalar field coupled to gravity.
     The ground state of a
     condensed cloud of charged bosons of mass $m$,
     interacting only gravitationally
     and having a total
 mass $M$ below a certain limit
 of the order $M_{\rm Pl}^2/m$,
 is a stable spherically symmetric configuration
 \cite{kau}
which is usually referred to as
 mini-soliton star \cite{fri2}
 or boson star \cite{jet}.
 The gravitational collapse of
boson stars is  prevented by
Heisenberg's uncertainty principle.
Seidel and Suen
have demonstrated that boson stars
may be formed through a dissipationless mechanism,
called gravitational cooling
\cite{sei}.
Boson stars have recently attracted much attention
as they may well be candidates for
nonbaryonic dark matter
\cite{mie}.

In this paper we study the thermodynamics of
boson stars
and its relation to
the Bose-Einstein condensation
in the framework of general relativity.
Some aspects of boson stars at finite temperature
have been studied
in Newtonian gravity \cite{ing}.
Our approach is similar in spirit to
the analysis of Q-ball
thermodynamics by
 Laine and Shaposhnikov
 \cite{lai}.
We will consider a canonical ensemble of charged
bosons and study its properties near the
critical temperature at which the Bose-Einstein condensate forms.

We organize the paper as follows:
In Section \ref{grand} we discuss  the grand canonical
ensemble of gravitating bosons
within a general-relativistic framework.
In Section \ref{canonical} the  free energy
is derived for a canonical system of gravitating bosons.
In Section \ref{numerical} we discuss the equations
that describe a boson star at finite temperature
and their numerical integration.
Numerical results are presented in Section \ref{results}.
Conclusions are drawn in
Section \ref{conclusions}.

\section{Grand canonical ensemble}
\label{grand}
In this section we derive the grand canonical partition
function for
a system of self-gravitating charged bosons
contained in a sphere of large radius $R$
in equilibrium at finite temperature $T=1/\beta$.
In a
 grand canonical ensemble we introduce the chemical
 potential $\mu$ associated to the conserved particle
 number $N$.
The partition function is given by
\begin{equation}
Z= {\rm Tr}\, e^{-\beta(H-\mu N)}=\int [dg][d\Phi]
 [d\Phi^*] e^{-S_{\rm g}-S_{\rm KG}},
\label{z}
\end{equation}
with the Euclidean actions
 $S_{\rm g}$
and $S_{\rm KG}$
 for the gravitational
and the Klein-Gordon  fields, respectively.
The gravitational part may be put in the form
\cite{gib,haw}
\begin{equation}
S_{\rm g}=-\frac{1}{16\pi}\int_{Y} d^4x \sqrt{g} {\cal R}
-\frac{1}{8\pi}\int_{\partial Y} d^3x \sqrt{h}(K-K_0),
\label{sg}
\end{equation}
where $\cal R$ is the curvature scalar,
$h$ is the determinant of the induced metric on the boundary,
and $K-K_0$ is the difference
in the trace of the second fundamental form of the
boundary $\partial Y$ in the metric $g$ and the flat metric.
The boundary  is a timelike tube of radius $R$
which is periodically identified in the imaginary time
direction with period $\beta$.
Thus, the functional integration assumes the periodicity
in imaginary time
and the asymptotic flatness of the metric fields.
The matter action is given by
\begin{equation}
S_{\rm KG}=\int_Y d^4x \sqrt{g}
(g^{\mu\nu}\partial_{\mu}\Phi^*
\partial_{\nu}\Phi
 +m^2 |\Phi|^2),
\label{skg}
\end{equation}
with the replacement of the Euclidean time derivative \cite{kap}
\begin{equation}
\frac{\partial}{\partial \tau} \rightarrow
\frac{\partial}{\partial \tau}\pm \mu ,
\label{dt}
\end{equation}
where the $+$ or $-$ sign is taken when the derivative
acts on $\Phi^*$ or $\Phi$,
respectively.
The path integral is taken over
asymptotically vanishing fields
which are periodic
in imaginary time $\tau$ with
period $\beta$.
The dominant contribution to the path integral comes from metrics
and matter fields, which are near the
classical  fields $g_{\mu\nu}$,
$\Phi^*$,
and $\Phi$.
The classical fields
extremize the action,
i.e., they are solutions to the classical field equations.
The background metric generated by the mass distribution is static,
spherically symmetric, and asymptotically flat, i.e.,
\begin{equation}
 ds^2 =e^{\nu(r)}dt^2-e^{\lambda(r)}dr^2
-r^2(d\vartheta^2+\sin ^2\vartheta \,
  d\varphi^2).
\label{ds}
\end{equation}
Substitution $t=-i\tau$ converts this into a positive
definite metric with the Euclidean signature,
\begin{equation}
 ds_{\rm E}^2 =e^{\nu(r)}d\tau^2+e^{\lambda(r)}dr^2
+r^2(d\vartheta^2+\sin ^2\vartheta \,
  d\varphi^2).
\label{dse}
\end{equation}
In order to extract the classical contribution,
we decompose $\Phi$ as
\begin{equation}
\Phi(x)=\phi(x)+\psi(x),
\label{Phi}
\end{equation}
where $\phi$ is a solution to the
Klein-Gordon equation
which we will call $\em condensate$
and $\psi$ describes  quantum and thermal
fluctuations around  $\phi$.
If we neglect quantum fluctuations
of the metric, the partition function factorizes as
\begin{equation}
Z=Z_{\rm g} Z_{\rm{cd}}
\int [d\psi][d\psi^*]e^{-S_{\rm KG}[\psi,\psi^*]},
\label{zz}
\end{equation}
where $Z_{\rm g}=e^{-S_{\rm g}}$ and
$Z_{\rm{cd}}=e^{-S_{\rm{cd}}}$ represent the saddle-point
 gravitational and condensate contributions, respectively.
The condensate  action
is given by
\begin{equation}
S_{\rm{cd}}=\int_Y d^4x \sqrt{g}
(g^{\mu\nu}\partial_{\mu}\phi^*
\partial_{\nu}\phi
 +m^2 |\phi|^2).
\label{s0}
\end{equation}
Variation of $S_{\rm{cd}}$ with respect to
$\phi^*$ yields a
Klein-Gordon equation
in which the time derivatives are replaced
by (\ref{dt}).
With the metric (\ref{dse}) and the usual ansatz for static
solutions
\begin{equation}
\phi(t,r) =\frac{1}{\sqrt{2}}e^{-i\omega't}\varphi(r)
=\frac{1}{\sqrt{2}}e^{-\omega'\tau}\varphi(r) \, ,
\label{phi}
\end{equation}
\begin{equation}
\phi^*(t,r)=\frac{1}{\sqrt{2}}e^{i\omega't}\varphi(r)
=\frac{1}{\sqrt{2}}e^{\omega'\tau}\varphi(r)  \, ,
\label{phi*}
\end{equation}
we obtain the static
Klein-Gordon equation
\begin{equation}
\frac{d^2\varphi}{dr^2}+
\frac{1}{2}\left(\frac{d\nu}{dr}-\frac{d\lambda}{dr}
+\frac{4}{r} \right) \frac{d\varphi}{dr} +
e^{\lambda}(\omega^2 e^{-\nu} -m^2)\varphi=0 \,,
\label{dphi}
\end{equation}
where $\omega=\omega'+\mu$.
Regular, asymptotically vanishing solutions to
equation (\ref{dphi}) coupled with Einstein field equations
are solitons which describe
the well-known boson stars
\cite{fri2,jet} at zero temperature.
Our aim  is to extend and analyze the corresponding solutions
at finite temperature.

Ignoring for the moment the periodicity condition,
the ansatz (\ref{phi},\ref{phi*}) gives the condensate
contribution to the partition function
\begin{equation}
\ln Z_{\rm{cd}}=\frac{1}{2}\beta
\int_{\Sigma} d^3x \sqrt{g}
\left[ g^{00}\omega^2\varphi^2
-g^{ii}(\partial_i \varphi)^2
 -m^2 \varphi^2 \right],
\label{lnz0}
\end{equation}
where $\Sigma$ is a spacelike hypersurface
that contains the condensate.
Using this expression
we find the net number of particles in the condensate
\begin{equation}
N_{\rm{cd}}=\frac{1}{\beta}\frac{\partial \ln Z_{\rm{cd}}}{
\partial \mu}
=\int_{\Sigma} d^3x \sqrt{g}
 g^{00}\omega\varphi^2 \, .
\label{n0}
\end{equation}
Alternatively, we define the particle number in a
covariant way as
\begin{equation}
N_{\rm{cd}}=
\int_{\Sigma} n_{\rm{cd}}
u^{\mu}d\Sigma_{\mu}
=\int_{\Sigma} d^3x \sqrt{g_{(3)}}\, n_{\rm{cd}} \,,
\label{n0s}
\end{equation}
where
$n_{\rm cd}$ is the particle-number density
in the condensate and
$g_{(3)}=\det (g_{ij})$, $i,j=1,2,3$.
Here we have used the
 fluid four-velocity
$u^{\mu}$
 the components of which in the comoving frame are
\begin{equation}
u^{\mu}=
\frac{
\delta^{\mu}_0
}{\sqrt{g_{00}}}\,  ;
 \;\;\;\;\;\;\;
u_{\mu}=
\frac{
g_{\mu 0}
}{\sqrt{g_{00}}}\,.
\label{u}
\end{equation}
Therefore, we identify the  particle-number density
due to the condensate as
\begin{equation}
n_{\rm{cd}}=
 \sqrt{g^{00}}
\omega\varphi^2.
\label{ncd}
\end{equation}
It may be easily verified that
the four-vector
$j^{\mu}=n_{\rm{cd}} u^{\mu}$
coincides with
the current
defined as
\begin{equation}
j^{\mu}=g^{\mu\nu}
(\phi\partial_{\mu}\phi^*
-\phi^*\partial_{\mu}\phi) .
\label{j}
\end{equation}

In order to satisfy
the periodicity condition,
i.e.,
$\phi(\beta)=\phi(0)$, we must set
$\omega'=0$.
This in turn implies $\omega=\mu$.
In other words,
as it was pointed out by
 Laine and Shaposhnikov
 \cite{lai},
if there exist  soliton solutions
with some values of $\omega$, these solutions are saddle points
of the Euclidean path integral at $\mu=\omega$.
In the absence of gravity,
a saddle-point solution is a homogeneous
Bose-Einstein condensate of a free relativistic
boson gas.

Next, we calculate the thermal
contribution to $Z$ starting from the
expression
\begin{equation}
Z_{\rm th}=
\int [d\psi][d\psi^*]e^{-S_{\rm KG}[\psi,\psi^*]},
\label{zth}
\end{equation}
with
\begin{equation}
S_{\rm KG} =
\int_Y d^4x \sqrt{g}
(g^{\mu\nu}\partial_{\mu}\psi^*
\partial_{\nu}\psi
 +m^2 |\psi|^2) .
\label{skg1}
\end{equation}
The metric in equilibrium is static, i.e.,
$g_{\mu\nu}$ is independent of $\tau$ and
$g_{0 i}=0$. Hence,
the  determinant of $g_{\mu\nu}$ factorizes
as
$g=g_{00}
g_{(3)}$.
By making use of the substitution
$\bar{\tau}=\tau \sqrt{g_{00}}$
we obtain
\begin{equation}
S_{\rm KG} =
\int d^3x \sqrt{g_{(3)}}
\int_0^{\bar{\beta}}d\bar{\tau}
{\cal L}(\bar{\tau},x) ,
\label{skg2}
\end{equation}
where
\begin{equation}
{\cal L}(\bar{\tau},x) =
(\partial_{\bar{\tau}}+\bar{\mu}) \psi^*
(\partial_{\bar{\tau}}-\bar{\mu}) \psi +
 g^{ij}\partial_i\psi^*
\partial_j\psi
 +m^2 |\psi|^2.
\label{l}
\end{equation}
Here we have defined the local chemical potential
and the inverse local temperature as
\begin{equation}
\bar{\mu} =\mu/\sqrt{g_{00}};
\;\;\;\;\;
\bar{\beta}=\beta\sqrt{g_{00}}.
\label{mu}
\end{equation}
These expressions are nothing but
the well-known Tolman conditions for chemical and
thermal  equilibrium in a gravitational field
\cite{tol,lan,ehl}.
Thus, the parameters $\beta$ and $\mu$
are interpreted as, respectively the values
of the inverse temperature and chemical potential at
infinity \cite{bar}.
In a close neighborhood $\Sigma_0$ of a space point
$X$
we introduce
local spatial coordinates $y= f_X (x)$ such that
\begin{equation}
 g^{ij}\frac{\partial\psi^*}{\partial x_i}
 \frac{\partial\psi}{\partial x_j}=
 \delta^{ij}\frac{\partial\psi^*_X}{\partial y_i}
 \frac{\partial\psi_X}{\partial y_j} ,
\label{gij}
\end{equation}
where
$\psi_X (y)=\psi (f^{-1}_X (y))$.
The Lagrangian (\ref{l}) is now
locally represented by the Euclidean flat-space Lagrangian
\begin{equation}
{\cal L}(\bar{\tau},X,y) =
 \left(\frac{\partial\psi_X^*}{\partial \bar{\tau}} +\bar{\mu}
 \psi^*_X\right)
 \left(\frac{\partial\psi_X}{\partial \bar{\tau}} -\bar{\mu}
 \psi_X \right)+
 \frac{\partial\psi^*_X}{\partial y^i}
 \frac{\partial\psi_X}{\partial y_i}
 +m^2 |\psi_X|^2 .
 \label{lt}
\end{equation}
By integrating
${\cal L}(\bar{\tau},X,y)$ over $y$
we define
\begin{equation}
{\cal L}(\bar{\tau},X) =
 \frac{1}{V_0}\int_{\Sigma_0} d^3y
{\cal L}(\bar{\tau},X,y) ,
 \label{ltx}
\end{equation}
where $V_0$ is the volume of
$\Sigma_0$.
Using the adiabatic approximation
${\cal L}(\bar{\tau},x) \approx
{\cal L}(\bar{\tau},X)$,
which holds for sufficiently small
$V_0$,
we obtain
\begin{equation}
S_{\rm KG} =
\int d^3X \sqrt{g_{(3)}(X)}
\int_0^{\bar{\beta}}d\bar{\tau}
 \frac{1}{V_0}\int_{\Sigma_0} d^3y
{\cal L}(\bar{\tau},X,y).
 \label{skg3}
\end{equation}
Now,
the partition function (\ref{zth})
may be factorized as
\begin{equation}
Z_{\rm th} =\prod_x Z_x  \, ,
 \label{zth1}
\end{equation}
where
\begin{equation}
Z_x =
 \int [d\psi]
[d\psi^*] \exp [
-\int d^4y\,d^4y'
 \psi^*(y)
 K_x(y,y')
 \psi(y')] ,
 \label{zx}
\end{equation}
with
\begin{equation}
 K_x(y,y')=
V_0^{-1}\sqrt{g_{(3)}(x)}\:
[(\partial_{\bar{\tau}}-\bar{\mu})^2 +\nabla_y^2)]
\delta^{(4)}(y-y') .
 \label{k}
\end{equation}
Assuming that $V_0$, although small, is still macroscopic
and that the thermodynamic limit may be applied to each $Z_x$,
 the standard functional integration technique \cite{kap}
 gives
\begin{eqnarray}
\ln Z_{\rm th}
\!&\!  =\! & \!
\int d^3x\, {\rm Tr} \ln K_x^{-1}    \nonumber \\
\!&\!  =\! & \!
-\int d^3x\,\sqrt{g_{(3)}}
 \int \frac{d^3q}{(2\pi)^3}
 [\ln (1-e^{-\bar{\beta}(E-\bar{\mu})})
 +\ln (1-e^{-\bar{\beta}(E+\bar{\mu})})] ,
 \label{lnzth}
\end{eqnarray}
where $E=\sqrt{q^2+m^2}$.
This expression may be regarded as a proper volume integral of
the local partition function
$\ln z(x)$
\begin{equation}
\ln z =
- \int \frac{d^3q}{(2\pi)^3}
 [\ln (1-e^{-\bar{\beta}(E-\bar{\mu})})
 +\ln (1-e^{-\bar{\beta}(E+\bar{\mu})})],
 \label{lnz}
\end{equation}
from which the
pressure,
energy density,
particle-number density,
and entropy density may be derived in the usual way:
\begin{equation}
p_{\rm th}=
\frac{1}{\bar{\beta}} \ln z,
 \label{pth}
\end{equation}
\begin{equation}
\rho_{\rm th}=
\left.
-\frac{\partial}{\partial
\bar{\beta}}
\ln z \right|_{\bar{\beta}\bar{\mu}} ,
 \label{rhoth}
\end{equation}
\begin{equation}
n_{\rm th}=
\frac{1}{\bar{\beta}}
\frac{\partial}{\partial
\bar{\mu}}
\ln z \, ,
 \label{nth}
\end{equation}
\begin{equation}
\sigma=
\bar{\beta}
(p_{\rm th} +
\rho_{\rm th} -
\bar{\mu} n_{\rm th}).
 \label{sigma}
\end{equation}
These expressions together with (\ref{lnz})
yield the well-known parametric representation of the
equation of state of a relativistic Bose
gas in curved space.
This equation of state is a special case of
a more general expression derived from
the relativistic kinetic theory
\cite{ehl}.

The gravitational part of the partition function
may be calculated
from (\ref{sg}) with help of Einstein field equations.
Using the result of Gibbons and Hawking
for the surface term
\cite{gib}, we obtain
\begin{equation}
\ln Z_{\rm g}=-\beta M+\int_Y d^4x \sqrt{g} \,
T_0^0  ,
\label{lnzg}
\end{equation}
where $M$ is the total mass and $T_{\mu}^{\nu}$ is
the energy-momentum tensor of the Klein-Gordon field
averaged with respect to the partition function (\ref{zz}).
The energy-momentum tensor of the Klein-Gordon field
can be obtained by variation of the
the Euclidean Klein-Gordon action with respect to the metric
\begin{equation}
 T_{\mu\nu}(\Phi)=
 -\frac{2}{\sqrt{g}}
 \frac{\delta S_{\rm KG}}{\delta g^{\mu\nu}}  .
\label{t}
\end{equation}
From (\ref{skg}) with (\ref{dt}) we find
\begin{equation}
T_{\mu\nu}(\Phi)=
-\partial_{\mu}\Phi^*
\partial_{\nu}\Phi
-\partial_{\nu}\Phi^*
\partial_{\mu}\Phi +
g_{\mu\nu}
(g^{\rho\sigma}\partial_{\rho}\Phi^*
\partial_{\sigma}\Phi
 +m^2 |\Phi|^2) .
\label{tmu}
\end{equation}
The averaged energy momentum tensor may be split up into
two parts:
\begin{equation}
 T_{\mu\nu}=
 T_{\mu\nu}(\varphi)+
 T_{{\rm th}\:\mu\nu},
\label{tmunu}
\end{equation}
where the first term on the right-hand side is
the classical part which comes from
the condensate and the
 second term
 represents
 the thermal and
 quantum fluctuations.
By making use of (\ref{phi}) and (\ref{phi*}) we
obtain the condensate contribution to $T^0_0$:
\begin{equation}
\rho_{\rm{cd}}=T_0^0(\varphi)=
\frac{1}{2}(g^{00}\omega^2\varphi^2
+g^{ii}(\partial_i \varphi)^2
 +m^2 \varphi^2) .
\label{rho0}
\end{equation}

 It may be shown that the thermal part is of the form
 which characterizes a perfect fluid:
\begin{equation}
 T_{{\rm th}\:\mu\nu}=
(\rho_{\rm th}+p_{\rm th})u_{\mu}u_{\nu}
-p_{\rm th}  g_{\mu\nu} \, ,
\label{tt}
\end{equation}
where the
thermal
pressure  and energy density
are given by
(\ref{pth}) and (\ref{rhoth}),
respectively.
Putting  the condensate (\ref{lnz0}),
the thermal (\ref{lnzth}), and
the gravitational (\ref{lnzg})
contributions together,
we find the total thermodynamical potential as
\begin{equation}
\Omega (\beta, \mu) =
-\frac{1}{\beta} \ln Z =
M-
\int_{\Sigma} d^3x \sqrt{g}
\left[ g^{00}\omega^2\varphi^2
+p_{\rm th}
+\rho_{\rm th}
 \right].
\label{omega}
\end{equation}
\section{Canonical ensemble}
\label{canonical}

Now consider
a system of self-gravitating charged bosons
with the net number of particles
 (number of particles minus number of antiparticles)
 $N$
contained in a two-dimensional sphere of large radius $R$
in equilibrium at finite temperature $T=1/\beta$.
In a
 canonical ensemble, instead of the chemical
 potential $\mu$ we fix  the particle number
which is the  sum of
the condensate
and thermal contributions.
Thus, a canonical ensemble is subject to
the constraint
\begin{equation}
\int_{\Sigma} d^3x \sqrt{g_{(3)}}
(n_{\rm{cd}}+n_{\rm th}) = N,
\label{n}
\end{equation}
where $n_{\rm{cd}}$ is given by (\ref{ncd})
and $n_{\rm th}$ by (\ref{nth}) with (\ref{lnz}).

The free energy of a canonical ensemble may
be derived from the grand-canonical partition function
with the help of the Legendre transform
\begin{equation}
F(\beta,N)=
\Omega (\beta, \mu) + \mu N.
\label{f}
\end{equation}
 The quantity $\mu$
 in this expression
 is an implicit
 function of $N$ and $T$, such that for given $N$ and $T$
 the constraint (\ref{n}) is satisfied
 {\cite{bilGRG}}.
 From (\ref{omega}) and (\ref{n}) with (\ref{ncd}) it follows
\begin{equation}
F =
M-
(\omega-\mu)\int_{\Sigma} d^3x \sqrt{g}
 g^{00}\omega\varphi^2
-\int_{\Sigma} d^3x \sqrt{g}
(p_{\rm th}
+\rho_{\rm th}-
\bar{\mu} n_{\rm th}).
\label{fm}
\end{equation}
If $\omega=\mu$,
the second term on the right-hand side vanishes
and the free energy may be expressed in the familiar form
\begin{equation}
F=
M-TS,
\label{fmt}
\end{equation}
where the total entropy $S$ is defined as a proper volume
integral
\begin{equation}
S=
\int_{\Sigma} \sigma
u^{\mu}d\Sigma_{\mu}
=\int_{\Sigma} d^3x \sqrt{g_{(3)}}\, \sigma,
\label{s}
\end{equation}
over the entropy density $\sigma$ given by
(\ref{sigma}).

\section{Numerical integration}
\label{numerical}

Given the temperature at infinity $T$, the radius of the
sphere $R$, and the particle number $N$,
we have to solve a set of self-consistency equations
consisting of the Klein-Gordon equation (\ref{dphi})
and Einstein field equations:
\begin{eqnarray}
\frac{d\nu}{dr} =8\pi
(p_{\rm{cd}}+ p_{\rm th})
r e^{\lambda} +\frac{1}{r}(e^{\lambda}-1) ,
\label{dnu} \\
\frac{d\lambda}{dr}=8\pi (\rho_{\rm{cd}} +\rho_{\rm th})
r e^{\lambda} -\frac{1}{r}(e^{\lambda}-1)  ,
\label{dlambda}
\end{eqnarray}
where the
thermal
pressure  and energy density
are given by
(\ref{pth}) and (\ref{rhoth}),
respectively,
the condensate density $\rho_{\rm{cd}}$ is given by (\ref{rho0}) and
the radial pressure of the condensate by
\begin{equation}
p_{\rm{cd}}=-T_r^r(\varphi)=
\frac{1}{2}(g^{00}\omega^2\varphi^2
+g^{ii}(\partial_i \varphi)^2
 -m^2 \varphi^2) .
 \label{trr}
\end{equation}
In addition, we impose the constraint (\ref{n})
which  fixes the chemical potential $\mu$ at infinity.
For numerical convenience, let us  introduce
the substitutions
\begin{equation}
e^{\nu(r)}=\frac{\omega^2}{m^2} \frac{1}{\chi(r)+1}  ,
\label{enu}
\end{equation}
\begin{equation}
e^{\lambda(r)}=\frac{1}{1-2{\cal M}(r)/r}  ,
\label{elambda}
\end{equation}
and new dimensionless
parameters
\begin{equation}
\alpha = \frac{\mu}{T}  ,
\; \;\;\;\; \gamma = \frac{\omega}{\mu} ,
\; \;\;\;\; \eta = \frac{m^2}{M_{\rm Pl}^2} ,
\label{alpha}
\end{equation}
where $M_{\rm{Pl}}=\sqrt{\hbar c/G}$
denotes the Planck mass.
Furthermore, we choose appropriate length and mass
 scales such that the length is measured in units
of $1/m$, and the mass in units of
 $M_{\rm{Pl}}^2/m$.
In this way,
all physical quantities appearing in our equations
become dimensionless and
equations (\ref{dphi}), (\ref{dnu}), and (\ref{dlambda})
now read
\begin{equation}
\frac{d^2\varphi}{dr^2}+\left( \frac{2{\cal M} -4\pi r^3(\tilde{\rho}
-\tilde{p})}{r-2{\cal M}} +2 \right) \frac{1}{r}
\frac{d\varphi}{dr} +\frac{r}{ r-2{\cal M}}\, \chi \,
\varphi =0 ,
\label{dphidr}
\end{equation}
\begin{equation}
\frac{d{\cal M}}{dr}=4\pi r^2 \tilde{\rho} ,
\label{dm}
\end{equation}
\begin{equation}
\frac{d\chi}{dr}=-2(\chi +1)\frac{{\cal M} +4\pi r^3
\tilde{p}}{r(r-2{\cal M})}.
\label{dchi}
\end{equation}
To these three equations we add
\begin{equation}
\frac{d{\cal N}}{dr}=4\pi r^2 (1-2{\cal M}/r)^{-1/2}
\tilde{n} ,
\label{dn}
\end{equation}
imposing the particle-number constraint as a condition at
the boundary:
\begin{equation}
{\cal N}(R)=N.
\label{nr}
\end{equation}
In equations (\ref{dphidr})-(\ref{dn}) we have used the abbreviations
\begin{equation}
\tilde{\rho}=\frac{\rho}{m^2 M_{\rm Pl}^2}=
\frac{1}{2} \left[
(\chi +2) \varphi^2 +\frac{r-2{\cal M}}{r}
\left(\frac{d\varphi}{dr}\right)^2 \right]
 +\eta \,
 \tilde{\rho}_{{\rm th}} \, ,
\label{rhotil}
\end{equation}
\begin{equation}
\tilde{p}=\frac{p}{m^2 M_{\rm Pl}^2}=
\frac{1}{2} \left[ \chi \varphi^2 +\frac{r-2{\cal M}}{r}
\left( \frac{d\varphi}{dr} \right)^2 \right]
 +\eta \,
 \tilde{p}_{{\rm th}}  \, ,
\label{ptil}
\end{equation}
\begin{equation}
\tilde{n}=\frac{n}{m M_{\rm Pl}^2}=
\sqrt{\chi+1} \varphi^2
 +\eta \,
 \tilde{n}_{{\rm th}} \, ,
\label{ntil}
\end{equation}
where the dimensionless quantities
$\tilde{\rho}_{{\rm th}}$,
$\tilde{p}_{{\rm th}}$, and
$\tilde{n}_{{\rm th}} $  are derived from
(\ref{lnz})-(\ref{nth}):
\begin{equation}
 \tilde{\rho}_{{\rm th}}=
 \frac{\rho_{{\rm th}}}{m^4}=
\frac{1}{2\pi^2}\int_{0}^{\infty} dy \, y^2 \sqrt{1+y^2} \, (n_+ +n_-),
\end{equation}
\begin{equation}
 \tilde{p}_{{\rm th}}=
 \frac{p_{{\rm th}}}{m^4}=
\frac{1}{6\pi^2} \int_{0}^{\infty} dy \, \frac{y^4}{\sqrt{1+y^2}} \,
(n_+ +n_-) ,
\end{equation}
\begin{equation}
 \tilde{n}_{{\rm th}}=
 \frac{n_{{\rm th}}}{m^3}=
\frac{1}{2\pi^2} \int_{0}^{\infty} dy \, y^2 \, (n_+ -n_-) \; ,
\end{equation}
\begin{equation}
n_{\pm}=\frac{1}{\exp \left[ \alpha \left( \gamma
\sqrt{(1+y^2)/(\chi+1)} \mp 1 \right) \right] -1} \, .
\label{npm}
\end{equation}
In equations (\ref{dphidr})-(\ref{npm})
the radial coordinate $r$,
the  enclosed mass $\cal M$,
and the particle number $ N$ are
measured in units
of $1/m$,
$m/\eta$,
$1/\eta$, respectively,
and the scalar field $\varphi$ in units
of $M_{\rm Pl}$.

It is evident from (\ref{rhotil})-(\ref{npm}) that
the thermal
contribution to the local quantities is suppressed by
the factor $\eta$, of
 the order of $10^{-38}$ for a boson mass
 of the order of 1 GeV.
 However, the contributions
 of the condensate and the thermal part to the global
 quantities, such as $M$, $N$, or $F$, may be of the same order of
 magnitude provided the volume of the system is sufficiently
 large.
 Indeed, the radius $R$ is much larger than
 the radial extent of the condensate, and,
 as we shall shortly demonstrate,
 the physically reasonable choice of $R$
makes
 the condensate and the thermal contributions
  comparable in magnitude.
  We shall also see that if we take the limit of
  vanishing $G$ keeping the central density fixed,
  we recover the free  boson gas
  in which the densities of the condensate and the gas
 are homogeneous
 and comparable in magnitude for the temperatures
 above zero and below the critical temperature.

Equations
(\ref{dphidr})-(\ref{dn}) should be integrated with
 boundary conditions dictated by physics.
 To avoid singularities at the origin, we take
\begin{equation}
{\cal{M}}(0)=0
\, ; \;\;\;\;\;
{\cal{N}}(0)=0
\, ; \;\;\;\;\;
\left. \frac{d\varphi}{dr}\right|_{r=0}=0.
\label{m0}
\end{equation}
The initial value of the scalar field
$\varphi_0\equiv\varphi(0)$
at $r=0$
is arbitrary but it
can be taken to be positive without loss
of generality.
 The metric field
 at the origin
$\chi_0\equiv\chi(0)$
is restricted to
\begin{equation}
-1<\chi_0\leq\gamma^2-1
\label{chi0}
\end{equation}
because of
the requirements that the metric should be positive
and that the Bose-Einstein distribution
(\ref{npm})
should be nonnegative everywhere.
However,
$\varphi_0$ and
$\chi_0$
 must be simultaneously tuned in order to satisfy
the constraint (\ref{nr})
and to provide the correct asymptotic behavior
$\varphi(r)\rightarrow 0$ as
$r\rightarrow \infty$.
The condition that
the field $\varphi$ should vanish asymptotically
is necessary  in order that the
 condensate  be either a soliton or
absent.
   A homogeneous condensate is excluded for
   $G\neq 0$ owing to our assumption
   that the metric should be asymptotically flat.

Once we have found a solution
 on the interval
$(0,R)$
for arbitrarily fixed parameters $\alpha$
and $\gamma$ with the correctly
chosen initial conditions
at $r=0$,
 we can determine the unknown physical parameters
 $\omega$, $\mu$, and $T$ from the boundary
 conditions at $r=R$.
We cut off the matter
from $R$ to infinity
and join the interior solution onto
the empty space Schwarzschild
exterior solution
\begin{equation}
\chi(r)=\frac{\omega^2}{m^2}
\left(1-\frac{2 M}{r}\right)^{-1}-1 .
\label{chir}
\end{equation}
Combining this equation with the numerically
found interior solution at $r=R$, we
obtain
$\omega$ which then, together with
 (\ref{alpha}),
 fixes  the chemical potential $\mu$
 and the temperature $T$ at infinity.

\begin{figure}[p]
\vbox{\vskip230pt
\includegraphics{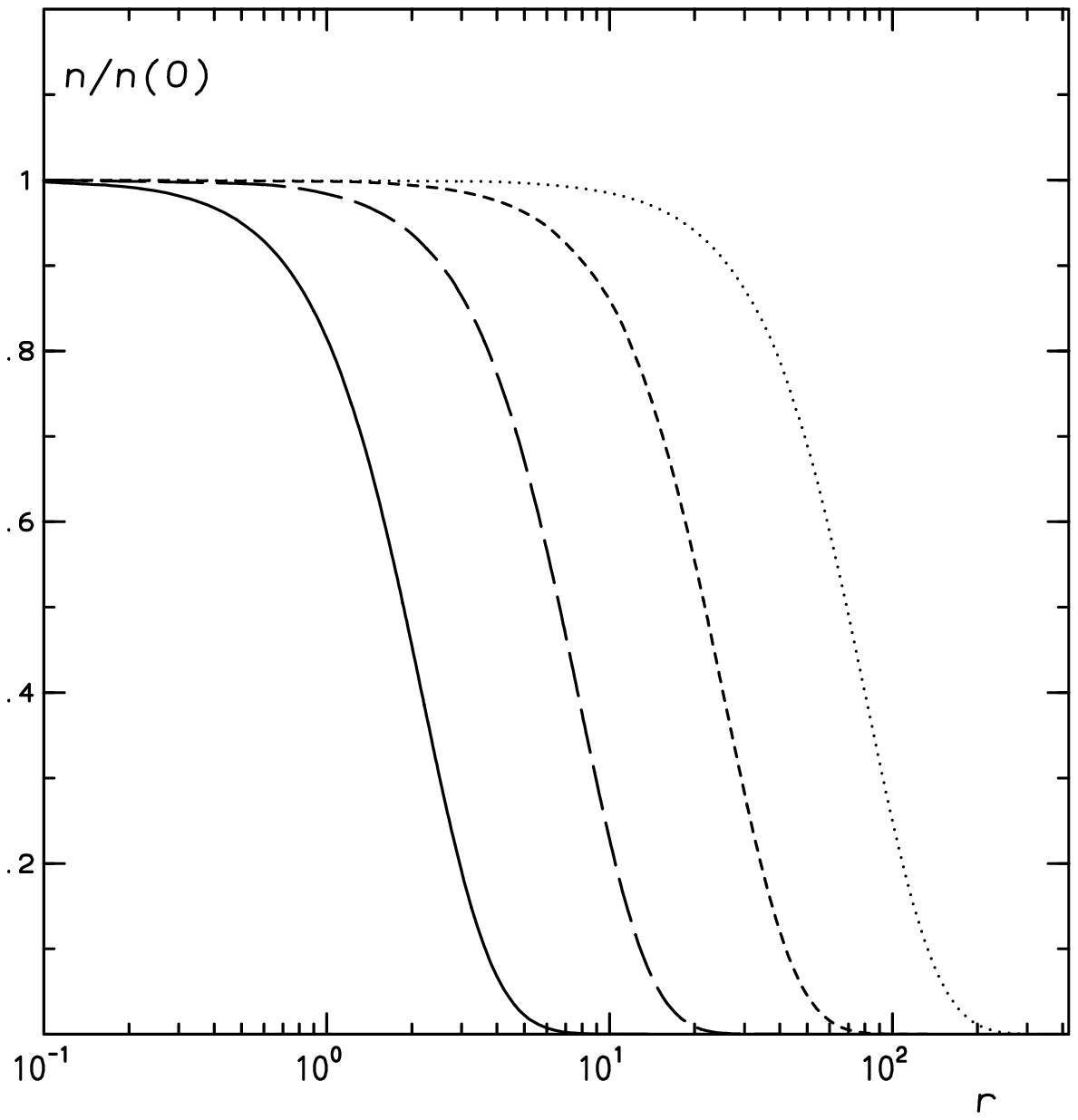}}
\caption{
Normalized particle-number density in
a boson star at $T=0$
for $\varphi_0$  equal to 0.0764 (full line),
0.01 (long dashed  line),
0.001 (short dashed  line),
and 0.0001 (dotted  line).
The radius $r$ is in units of $m^{-1}$}
\label{nn_}
\end{figure}

Consider first the  condensate
at zero temperature.
We know that
 a zero-node
 soliton solution exist  for any
 particle number  $N$ below
 $N_{\rm max}=0.653$
 \cite{fri2}.
 This maximal value,
 known as the {\em Kaup limit}
 \cite{kau},
 corresponds to
 the initial values
 $\varphi_0=0.0764 $, $\chi_0=0.545$,
 and the frequency $\omega=0.853 m$.
 Lower $\varphi_0$ lead to lower $N$
 and larger $\omega$.
 In the limit
$ \varphi_0\rightarrow 0$,
 one approaches the Newtonian regime
 in which
$\omega$ approaches the
 limiting value $m$.
 It may be easily shown that
 in the Newtonian limit there exist
 a mass-radius relationship of the form
\begin{equation}
M R_0={\rm const}
\frac{M_{\rm Pl}^2}{m^2} ,
\label{mr}
\end{equation}
where the radius $R_0$ of the boson star may
be conveniently defined, e.g.,
as the radius of a ball of mass $M$
with constant density  equal
to the central density
$\rho(0)=(1+\chi_0/2)\varphi_0^2 m^2 M_{\rm Pl}^2$.
Other definitions are possible \cite{fri2}
and lead to the same qualitative
conclusions.
Relation (\ref{mr}) enables us to analyze
the fate of the boson star in the  limit
$G\rightarrow 0$.
In that limit, we keep a certain physical quantity fixed,
e.g., the  particle number $N$, or the central density $\rho(0)$.
In both cases, the radius $R_0$ increases with decreasing $G$
and becomes infinite for $G=0$.
For example, if $\rho(0)$ is kept fixed,
$R_0$ grows as
$G^{-1/4}$
as a consequence of
(\ref{mr}).
This situation is illustrated in Fig.\ \ref{nn_}
where we plot the number-density profile of the
boson star in the ground state
for several initial values $\varphi_0$.
The density profile is in the form of a plateau
the length of which roughly measures the
radius of the boson star.
With decreasing $\varphi_0$, which corresponds to decreasing $G$,
the plateau widens and becomes more pronounced.
In the limit $G\rightarrow 0$, the boson star
becomes an ordinary homogeneous condensate,
with $\omega=m$.

We now turn to the study of nonzero temperature.
As discussed in section \ref{grand},
thermodynamic consistency
strictly requires $\gamma=1$ which
 implies $\chi_0\leq 0$
 owing to (\ref{chi0}).
It may be shown
that if
 $\chi_0\leq 0$
 and $\varphi_0>0$,
 then the function
 $\varphi(r)$ is monotonically increasing
 with $r$.
 To see this, note that near the origin
 the first derivative of
 $\varphi$
 behaves as
 $\varphi'\approx
 - \varphi_0\chi_0 r/3$,
 and hence,
 $\varphi$ increases
 for small $r$.
 Furthermore,
 at every point $r>0$
 where
 $\varphi'(r)=0$,
 the second derivative
 $\varphi''(r)$ is negative
 as a consequence of equation (\ref{dphidr}).
 This implies that the function
 $\varphi$
 never reaches a maximum.
Thus,
equation (\ref{dphidr}) with
(\ref{dm})-(\ref{dchi}) possesses no
soliton solution if
 $\chi_0\leq 0$,
 and at any temperature $T>0$,
  the only physically
 acceptable solution,
 would be
 the trivial one $\varphi\equiv 0$.
It seems that, in contrast to the self-gravitating
 fermion gas
 \cite{bilEPJ},
 a self-gravitating boson gas
 does not possess a nontrivial
 limiting configuration as $T\rightarrow 0$.
 Moreover, if we hold $N$ fixed and decrease the temperature
 sticking to $\gamma=1$, i.e.,
 $\varphi\equiv0$,
 $\mu$ will increase until
 we  reach  a  limiting temperature $T_{\rm{c}}$
 for which $\bar{\mu}(0)=m$.
 Further decrease of $T$
with fixed $N$ is no longer possible
 unless a condensation process takes place.
 If there were no gravity,
 we would decrease the temperature below $T_{\rm{c}}$ keeping $\mu=m$.
As a result, the thermal density  of particles would decrease
and the density of particles in the ground state
would adequately increase forming
a homogeneous Bose-Einstein condensate
$\varphi= {\rm const}$.
In that process, the total number of particles would remain fixed.

In the presence of gravity,
the formation of a homogeneous condensate is not
possible
since $\varphi={\rm const}$  is not a solution to
(\ref{dphidr}).
However,
a condensation
process will  take place
if below $ T_{\rm{c}}$ the formation of a soliton is
made possible
by allowing $\chi_0>0$.
This implies that the chemical potential
at infinity $\mu$ is less than $\omega$,
although the local chemical potential at the origin
$\bar{\mu}(0)$  stays equal to $m$.
The reason for this apparent thermodynamic
inconsistency lies in the fact that general relativity
does not allow  configurations with
asymptotically constant matter density.
Of course, in the limit
 $G\rightarrow 0$, the soliton solution would
become an ordinary homogeneous condensate,
with $\mu=\omega=m$,
 as in the $T=0$ case discussed above.
 In numerical calculations we shall stick to
 the condition
$\bar{\mu}(0)=m$ below $T_{\rm c}$.
This condition,
which prevents the
Bose-Einstein distribution from becoming negative,
fixes
$\gamma=\sqrt{\chi_0+1}$ for  $\chi_0>0$,
and $\gamma=1$ for $\chi_0 \leq 0$.

To proceed with the numerics, we have to specify the radius $R$.
Although
 $R$ is an arbitrary parameter,
the following physical considerations will give us
its preferable order of magnitude.
First, the size of the system is clearly much larger than
the size that it would assume at zero temperature,
i.e., $R$ must be larger than the radius of
a boson star $R_0\sim 1/m$.
Second, it should be smaller or at most
of the order of
the size it would take as a gas at high temperatures.
To estimate the natural size of the purely gaseous phase,
note that if the condensate vanishes,
the quantities ${\cal M}$ and $r$ may be rescaled
so that the dimensionless
field equations
remain in the same form as in
(\ref{dm})-(\ref{dchi})
with
$\tilde{\rho}$ and
$\tilde{p}$ replaced by
$\tilde{\rho}_{\rm th}$ and
$\tilde{p}_{\rm th}$, respectively,
and with $r$ and ${\cal M}$ measured in units of
$M_{\rm Pl}/m^2$ and
$M_{\rm Pl}^3/m^2$, respectively.
Thus, the natural size of a purely gaseous phase is
of the order
$M_{\rm Pl}/m^2$ and therefore we expect the radius of our system
to be in the range $1/m \ll R$
 \raisebox{-1.5mm}{$\stackrel{\textstyle <}{\sim}$}
$M_{\rm Pl}/m^2$.

Now suppose that
at an early stage of the Universe evolution, when the
temperature is high, say
 $T$ \raisebox{-1.5mm}{$\stackrel{\textstyle >}{\sim}$} $m$,
a gravitating boson ensemble exists in a purely gaseous
phase with the particle-number density of the order
$n_{\rm th}\sim m^3$ within a volume
of large radius $R_{\rm gas}$.
During the cooling down to temperatures
$T\ll m$ the
condensation will take place, in which a number of boson stars
will  be formed.
During this evolution
the particle number
 $N \sim
 R_{\rm gas}^3m^3$
is conserved.
Since
a typical particle number in a boson star is of the order
$1/\eta$,
the number of boson stars formed
in the condensation process is
approximately
 $n\sim\eta N \sim
 \eta R_{\rm gas}^3 m^3 $
Therefore, the volume occupied by each of the
boson stars is of the order $R_{{\rm gas}}^3/n$
and the corresponding radius of the order
\begin{equation}
R\sim
 \frac{R_{\rm gas}}{
n^{1/3}}
\sim \frac{1}{\eta^{1/3}m}.
\label{r}
\end{equation}
This quantity is much larger than the radius $R_0$ of the boson star
itself, so the condensate
will occupy only a small portion of the volume
 $V\sim R^3$.
At $T\sim m$, the thermal contribution to the particle number
is $N_{\rm th}\sim R^3n_{\rm th} \sim 1/\eta$.
Therefore, the contributions to the total charge
of the thermal part and of the condensate
are of the same order of magnitude despite the apparent
incommensurability of the two densities in (\ref{ntil}).

\section{Results and discussion}
\label{results}

To integrate  equations (\ref{dphidr})-(\ref{dn})
numerically,
we take
the radius of the system
as
$R=\eta^{-1/3} m^{-1}$,
in accordance with
(\ref{r}),
and, for definiteness,
we take the mass of the boson as
$m=1{\rm GeV}$.
The choice of $m$ will
practically not affect our results
as long as $m\ll M_{\rm Pl}$.

\begin{figure}[p]
\vbox{\vskip230pt
\includegraphics{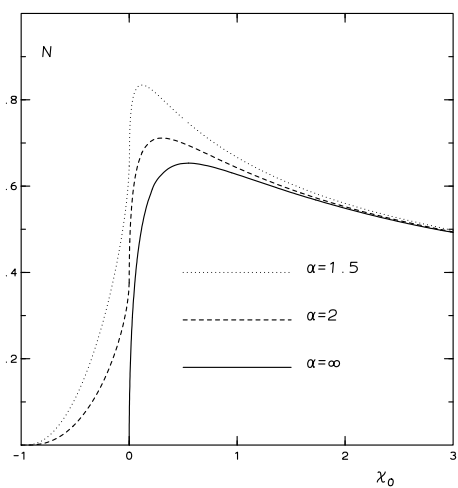}}
\caption{
Particle number $N$
(in units of
$M_{\rm  Pl}^2/m^2$),
versus $\chi_0$ for various $\alpha$.
}
\label{n_}
\end{figure}

For given $\alpha$, the initial values
$\chi_0$ and $\varphi_0$
that yield the ground-state (zero-node) solution
with the required particle number
are not always uniquely determined.
The reason is that, owing to general-relativistic
effects, there exists a finite $\chi_0$ for which
$N$ and $M$ reach a maximum.
This limiting configuration is similar
to the Oppenheimer-Volkoff limit
for the degenerate fermion stars.
Further increase of $\chi_0$ makes $N$
oscillate about a nonzero limiting value $N_{\infty}$
that corresponds to the infinite central density.
In Fig.\ \ref{n_} we plot the particle number $N$
versus $\chi_0$ for various $\alpha$, including
$\alpha=\infty$ which corresponds to zero temperature.
It is clear that for each $\alpha$
there will be at least two configurations
with the same $N$, if $N$ is slightly below the maximum.
However, the configurations represented by the points
on the right of the  maximum on each curve
are thermodynamically unstable since their
free energy,
as we shall shortly see,
is larger than the free energy of the corresponding
configurations on the left of the maximum.
Therefore, if we restrict our consideration
to stable configurations, the quantities
$\chi_0$ and $\varphi_0$
will be uniquely determined.

\begin{figure}[p]
\vbox{\vskip230pt
\includegraphics{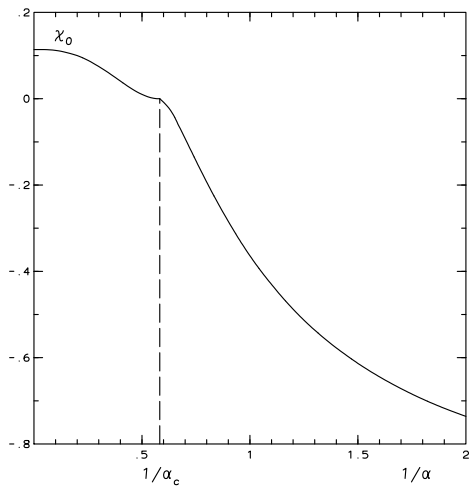}}
\caption{Initial value of the metric field $\chi_0$
that yields the zero-node soliton solution with $N=0.5
M_{\rm  Pl}^2/m^2$,
 within a sphere of radius
$R=(M_{{\rm Pl}}/m)^{2/3} m^{-1}$, versus
$1/\alpha$.}
\label{chi0_}
\end{figure}

\begin{figure}[p]
\vbox{\vskip230pt
\includegraphics{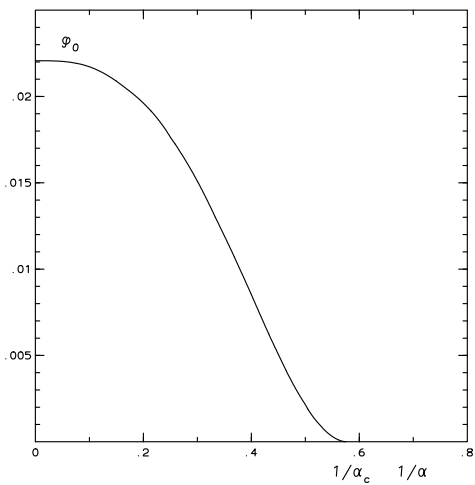}}
\caption{Initial value $\varphi_0$ as a function of $1/ \alpha$
for   $N$ and $R$ as in Fig.\ \protect{\ref{chi0_}}.}
\label{phi0_}
\end{figure}

Fixing
$N=0.5$, which is below the maximum of the
zero temperature curve, we now plot the initial
values
 $\chi_0$ and $\varphi_0$
 as functions of inverse $\alpha$
in Figs.\ \ref{chi0_} and \ref{phi0_}, respectively.
At the  critical point $1/\alpha_{\rm c}=T_{\rm c}/m=0.582$,
$\chi_0$ changes the sign and $\varphi_0$ vanishes.
Thus, the configurations  for which $\alpha < \alpha_{\rm c}$
are purely thermal and those with
 $\alpha > \alpha_{\rm c}$ are mixed.
For each $\alpha$ we can find $\omega$, $\mu$, and
$T$ by making use of equations (\ref{alpha}) and
(\ref{chir}).

\begin{figure}[p]
\vbox{\vskip230pt
\includegraphics{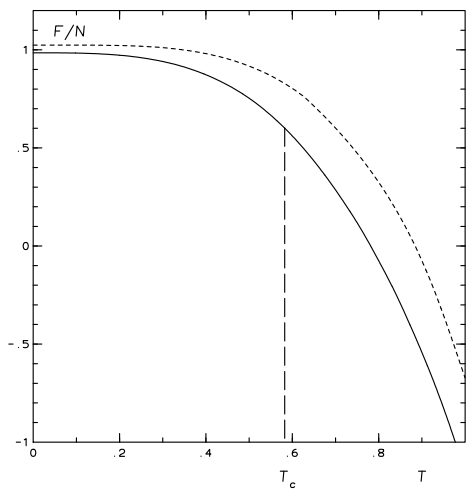}}
\caption{Free energy per particle
$F/N$ (in units of $m$) versus temperature T (in units of $m$)
for   $N$ and $R$ as in Fig.\ \protect{\ref{chi0_}}.}
\label{f_}
\end{figure}

\begin{figure}[p]
\vbox{\vskip230pt
\includegraphics{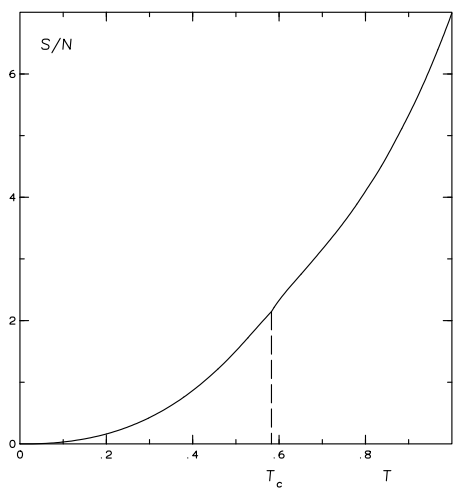}}
\caption{Entropy per particle versus temperature (in units of $m$)
for   $N$ and $R$ as in Fig.\ \protect{\ref{chi0_}}.}
\label{s_}
\end{figure}

\begin{figure}[p]
\vbox{\vskip230pt
\includegraphics{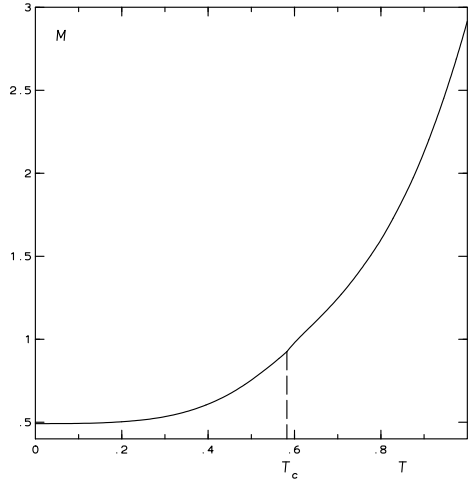}}
\caption{Total mass (in units of $M_{{\rm Pl}}^2/m$)
versus temperature (in units of $m$)
for   $N$ and $R$ as in Fig.\ \protect{\ref{chi0_}}.}
\label{m_}
\end{figure}

In Fig.\ \ref{f_} we plot the free energy per
particle,
calculated using (\ref{fmt}) and (\ref{s}),
as a function of
temperature.
The full line represents the free energy  of
the configurations corresponding to
the points on the left of the maxima in
Fig.\ \ref{n_},
whereas
the dashed line represents the free energy
corresponding to the
points on the right of the maxima.
As expected, the latter free energy
is everywhere larger than the
former, hence,
the configurations it represents are thermodynamically unstable.
In what follows we  restrict our attention to
stable configurations only.
The critical temperature
$T_{\rm c}= 0.582\; m$
is indicated by a vertical line.
In Figs.\ \ref{s_} and \ref{m_}
the entropy per particle and the total mass
are plotted
as functions of temperature.
Since $S$ and $M$ are both continuous functions
at $T_{\rm c}$, we conclude that the process of
condensation is a second-order phase transition
with the naturally defined order parameter $\varphi_0$
plotted in Fig.\ \ref{phi0_}.

   In order to facilitate the physical understanding of our
   results, let us mention the properties of the
   particular soliton configuration in physical units.
   The radius of the soliton  at $T=0$  is a few times larger than
   $(cm/\hbar)^{-1}$, so for $m=1$ GeV it is
   approximately $1.2\times 10^{-12}$ cm,
   whereas  the radius of the whole star at $T\neq 0$  is 1.04 cm.
   The total mass of the soliton  at $T=0$  is
   $M=1.3\times 10^{11}$ kg, the particle number
   $N=7.5\times 10^{37}$, and
   the critical temperature  $T_c =0.582$ GeV.

At this stage it is instructive to compare our results with
the approximate analytical result of a corresponding
free-boson system.
Taking the lowest contributions in the high-temperature
expansion, one can find simple analytical expressions that describe
properties of the free boson gas \cite{hab2}.
In particular, the
critical temperature is given by
\begin{equation}
T_{\rm c}=\left( \frac{3N}{mV} \right) ^{1/2}  .
\label{tc}
\end{equation}
Using $V=4\pi R^3/3$ we find
\begin{equation}
\frac{T_{\rm{c}}}{m}=\frac{3}{2\sqrt{\pi}} R^{-3/2}N^{1/2},
\label{tcm}
\end{equation}
with $R$ measured in units of
 $\eta^{-1/3}m^{-1}$ and $N$ in units of
 $\eta^{-1}$.
 Within the same approximation,
 the temperature dependence of the chemical potential
 is given by
\begin{equation}
\frac{\mu}{m}=\left( \frac{T_{\rm{c}}}{T} \right)^2 \; ,
\label{mum}
\end{equation}
for $T\geq T_{\rm{c}} \gg m$, and $\mu=m$
for temperatures $T\leq T_{\rm c}$.

\begin{figure}[p]
\vbox{\vskip230pt
\includegraphics{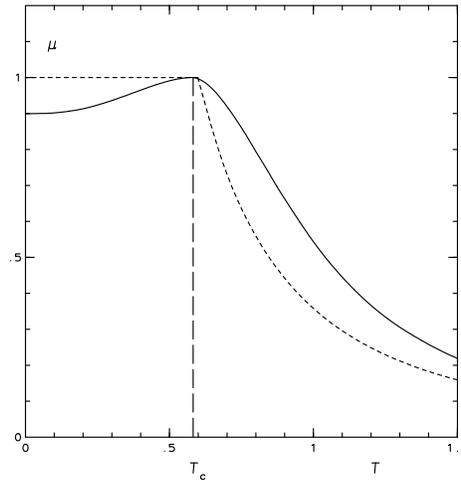}}
\caption{Chemical potential $\mu$ (in units of $m$) at $r=\infty$ as a
function of temperature (in units of $m$) for gravitating bosons (solid line)
and for free bosons in the high-temperature approximation
(dashed line)
for   $N$ and $R$ as in Fig.\ \protect{\ref{chi0_}}.}
\label{mu_}
\end{figure}

\begin{figure}[p]
\vbox{\vskip230pt
\includegraphics{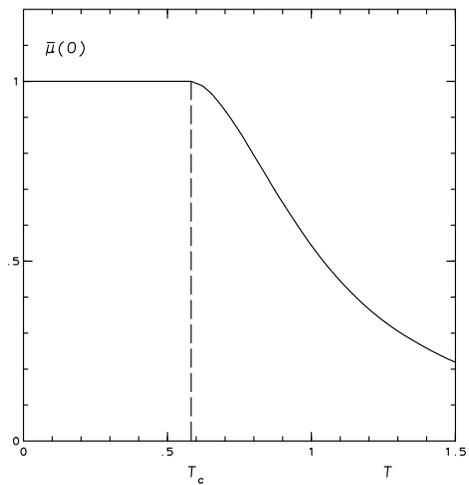}}
\caption{Chemical potential $\bar{\mu}$ (in units of $m$) at $r=0$
as a function of temperature (in units of $m$)
for   $N$ and $R$ as in Fig.\ \protect{\ref{chi0_}}.}
\label{mu0_}
\end{figure}

In Figs.\ \ref{mu_} and \ref{mu0_}
we plot the chemical potential $\mu$ at infinity
and the chemical potential $\bar{\mu}(0)$
at  the origin,
respectively, as functions
of temperature.
For comparison,
in Fig.\ \ref{mu_} we also give
the discussed approximate analytical result
for a free boson system.
It is worth noting the following
two effects.
First, the critical temperature is almost
exactly equal to that obtained for
a free relativistic Bose system.
Second, the functions describing the temperature dependence of
 $\mu$ and $\bar{\mu}(0)$
for $T>T_{\rm{c}}$
are almost exactly the same.
The reason for both effects is
that $g_{00}(r)$ in a purely gaseous phase
is practically constant on the interval  $(0,R)$.
The departure of the analytical result
for a free boson system
from the numerical result for a gravitating boson system
is a general-relativistic effect
below $T_{\rm c}$,
and
above $T_{\rm c}$ is just due to the high-temperature
approximation.

Our results depicted in Figs.\ \ref{n_}-\ref{mu0_}
depend, of course, on the choice of $R$ and $N$.
However, as in the case of a free gas, the critical temperature
should not depend on $R$ nor on $N$ if
the average density is kept fixed.
We have checked that, if we keep the ratio
$N/R^3$ fixed,
our $T_{\rm c}$ remains constant to very high
accuracy
with $R$ varying by several orders of magnitude.

So far, we have studied the condensation process
assuming $m\ll M_{\rm Pl}$, i.e., $\eta\ll 1$.
Now, let us investigate
 an extreme general-relativistic case
when the value of the boson mass is a sizable
fraction of the Planck mass, say
  $m=0.1 M_{\rm Pl}$, yielding $\eta=0.01$.
  In this case,
 the natural length scale
 $\eta^{-1/2}m^{-1}$
 of the purely thermal phase
 becomes comparable with the natural size
 $\eta^{-1/3}m^{-1}$ of the mixed phase,
 and we expect large gravitational effects.

\begin{figure}[p]
\vbox{\vskip230pt
\includegraphics{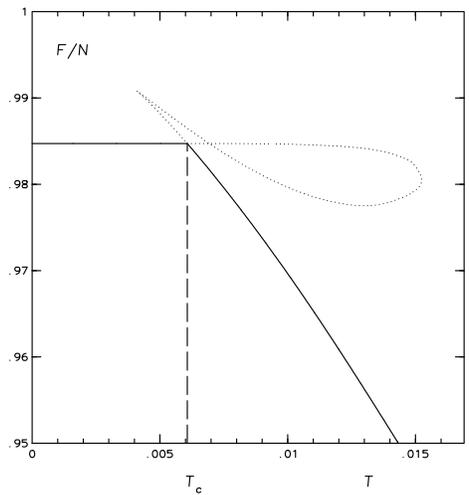}}
\caption{Free energy per particle (in units of $m$)
versus temperature (in units of $m$)
for $\eta\equiv
m^2/M_{{\rm Pl}}^2=0.01$
and  $N=0.5
M_{{\rm Pl}}^2/m^2$.}
\label{fn_}
\end{figure}

\begin{figure}[p]
\vbox{\vskip230pt
\includegraphics{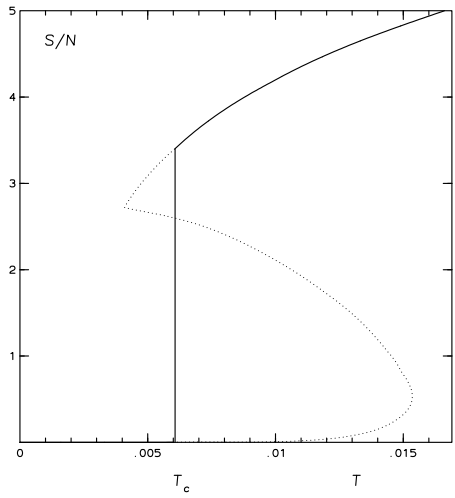}}
\caption{Entropy per particle versus temperature (in units of $m$)
for $\eta\equiv
m^2/M_{{\rm Pl}}^2=0.01$
and  $N=0.5
M_{{\rm Pl}}^2/m^2$.}
\label{sn_}
\end{figure}

The calculations  are presented
in Figs.\ \ref{fn_} and \ref{sn_} in which, respectively,
the free energy and entropy are shown as functions of
temperature for $\eta=0.01$ and $R=10\eta^{-1/2}m^{-1}$.
 The part of the curve in Fig.\ \ref{fn_}
that starts from $T=0$, makes the loop,
 and ends at the sharp cusp, represents the mixed
 soliton-gas phase.
 The rest of the curve,
which can be continuously extended to high temperatures,
 represents the purely gaseous phase.
In the temperature interval
 $T=(0.003-0.015) m$
there are
three distinct solutions of which
 only two are physical,
 namely, those
 for which the free energy
  assumes a minimum.
  The cusp corresponds to the naive transition point.
  However, the actual transition takes place at
  the temperature $T_{\rm c}$,
  where the free energy of the mixed phase and that of the
gas become equal.
Hence, it is a first-order phase transition.
The dotted curves in Figs.\
 \ref{fn_} and \ref{sn_}
 represent the physically unstable solution.
In our example,
the transition temperature is
$T_c=0.00607 m$,
as indicated in the plots by the dashed line.
The latent heat per particle released
during the phase transition is given by the
entropy difference at the point of
discontinuity
\begin{equation}
\frac{\Delta M}{N}=
\frac{\Delta S}{N}T_{\rm c}=0.0206m.
\end{equation}
This phase transition is similar to the gravitational
phase transition in self-gravitating fermionic systems
\cite{bilEPJ}, with one important
distinction:
in a gravitating bosonic system
the
phase transition
is first  order only in the extreme
general-relativistic regime,
whereas
in gravitating fermionic systems
it remains first order
even in the Newtonian regime \cite{mes,bilPL}.
\section{Conclusions}
\label{conclusions}
In this work, we have studied a canonical system  of
self-gravitating bosons at finite temperature
in a general-relativistic framework.
We have numerically solved the
system of self-consistency equations
consisting of a Klein-Gordon equation coupled to
Einstein field equations and the equation of state for
a gravitating boson gas.
 We have
  investigated
the circumstances under which
this system undergoes a
Bose-Einstein condensation.
This condensation is quite distinct from
the usual one in that it involves the formation of
a soliton with the spherically symmetric
matter distribution concentrated
around the origin,
as opposed to the usual spatially homogeneous
Bose-Einstein condensate.
In the $T\rightarrow 0$ limit, the soliton becomes
a mini boson star.
The condensation begins with a second order phase transition
at the critical temperature $T_{\rm c}$.
The system exists in two phases:
a gaseous phase above
 $T_{\rm c}$
and the mixed soliton-gas phase below $T_{c}$.
The critical temperature is approximately
proportional to the square root of
the average particle-number density
and is very close the the corresponding
critical temperature of a free boson gas.

General relativistic effects become important
when the boson mass is a few orders of magnitude
away from the Planck mass.
In that case,
the condensation begins with a first-order
phase transition that
qualitatively resembles the gravitational
phase transition of fermionic matter.

\section*{Acknowledgments}
 This work  was supported by
 the Ministry of Science and Technology of the
 Republic of Croatia under Contract
 No. 00980102.

\end{document}